\documentclass[twocolumn,showpacs,aps,pre,
groupedaddress,amssymb,amsmath]{revtex4}

\usepackage{graphicx}
\usepackage{longtable}

\begin{document}

\title{Impact of noise on domain growth in electroconvection}
\author{Mary Griffith and Michael Dennin}
\affiliation{Department of Physics and Astronomy, University of
California at Irvine, Irvine, California 92697-4575}

\date{\today}

\begin{abstract}

The growth and ordering of striped domains has recently received
renewed attention due in part to experimental studies in diblock
copolymers and electroconvection. One surprising result has been
the relative slow dynamics associated with the growth of striped
domains. One potential source of the slow dynamics is the pinning
of defects in the periodic potential of the stripes. Of interest
is whether or not external noise will have a significant impact on
the domain ordering, perhaps by reducing the pinning and
increasing the rate of ordering. In contrast, we present
experiments using electroconvection in which we show that a
particular type of external noise decreases the rate of domain
ordering.

\end{abstract}

\maketitle

The study of domain ordering after a sudden change to a system (or
a quench) has implications for a wide range of applications and
fields, from processing binary mixtures of fluids to the
organization of crystalline domains in a solid \cite{REV}. Domain
ordering occurs when different spatial regions of a system are in
different states, and the size of these regions change with time.
A common method to generate such a situation is to take a uniform
system and subject it to a sudden change in external parameters
such that the system can now exist in two or more degenerate
states. Regions form that select from the possible states,
creating an inhomogeneous system that proceeds to order. Our
understanding of the ordering process focuses on the case when the
system is spatially uniform within each domain. In this case, one
can generally understand the coarsening, or domain ordering, by
considering the motion of the topological defects in the system
\cite{REV}. The situation is  less clear when the domains
themselves contain spatial structure. In this regard, domains of
stripes have received significant attention
\cite{EVG92,CM95,CB98,HAC00,BV01,HCSH02,QM03,QM04,PD01,TS99,TNS02,B04}.

Stripes, or more generally patterns, occur in a wide range of
systems \cite{CH93,GL99}, including convecting fluids, animal
coats, polymer melts, and ferromagnets. Stripes occur both as an
equilibrium state of the system, such as in diblock copolymers,
and as a result of external driving, as in convection in fluids.
Theoretical studies of the ordering of striped domains
\cite{EVG92,CM95,CB98,BV01,QM03,QM04,TS99,TNS02,B04} have focused
on studies of model equations, such as the Swift-Hohenberg
equation. In general, simulations find that the growth of striped
domains occurs slower than might be expected from our knowledge of
the growth of uniform systems. This is typically characterized by
the growth exponent. For sufficiently late times, it is postulated
that the length scale of domains in these systems scales as a
power of time, usually referred to as the growth exponent. For
uniform domains in systems approaching an equilibrium state, it is
known that the exponents are $1/2$ if the order parameter is not
conserved and $1/3$ for a conserved order parameter \cite{REV}.
For striped systems, growth exponents of various values are
reported, including $1/3$, $1/4$, and $1/5$. There is evidence
that for sufficiently large quenches, the system becomes glassy,
and scaling breaks down \cite{BV01,B04}. Of particular interest to
the work in this paper are simulations that focus on anisotropic
systems \cite{B04} and recent simulations that focus on the impact
of noise on the coarsening of stripes \cite{TS99,TNS02}. This will
be discussed in more detail later.

The ordering of stripe domains has been studied experimentally
both for the diblock copolymer case \cite{HAC00,HCSH02} and the
electroconvection in a nematic liquid crystal
\cite{PD01,KFD04,KID04}. The work with diblock colpolymers
strongly suggests that topological defects play a central role in
the ordering of stripe domains \cite{HCSH02}. The work in
electroconvection is interesting because the system is an example
of coarsening in an anisotropic, driven system. The work in this
paper focuses on domain coarsening in this system.

Nematic liquid crystals are fluids in which the molecules exhibit
long-range orientational order \cite{GP93}. The average axis along
which the molecules are aligned is referred to as the director. By
proper treatment of the boundaries, samples with uniform director
alignment can be prepared. When an electric voltage is applied to
such a sample, there is a critical voltage at which convective
flow of the fluid occurs. There is a corresponding periodic
variation of the director field. This pattern is known as
electroconvection \cite{BZK88,KP95,RWTSS89}. When the axis of the
convection rolls forms a nonzero angle with the undistorted
director field, the pattern is referred to as oblique rolls. This
is a degenerate state because patterns with an angle $\theta$ (zig
rolls) and $-\theta$ (zag rolls) are degenerate. By applying a
sudden increase in voltage from below to above the critical value,
a pattern of zig and zag domains is created. This pattern
coarsens, or orders, in time. The ordering of the system is
consistent with power law growth in time \cite{PD01}, but the
growth is anisotropic, occurring at different rates parallel and
perpendicular to the director \cite{KFD04}.

Because electroconvection is driven electrically, it is a
relatively easy system for studies of the impact of noise.
Simulations of an anisotropic Swift-Hohenberg model \cite{B04}
suggest that the pinning of defects is relevant to the coarsening
of the domains in electroconvection. Simulations for isotropic
systems suggest that noise can alter the dynamics of the domain
growth by changing the potential in which the defects move,
effectively acting as a thermal bath \cite{TS99,TNS02}. Though
these simulations were performed for isotropic systems, there is
no obvious reason that a similar phenomenon would not occur in
electroconvection.

In considering the impact of noise on the domain growth, one can
imagine different types of noise. The two classic cases are
additive noise and multiplicative noise. These are best defined in
the context of an amplitude equation formulation (or envelope
equation) in which the fast variation (the period corresponding to
the fundamental pattern) is removed and only long wavelength
changes in the pattern are studied. In this formalism, additive
noise is described by the addition of a noise term to the equation
that is not directly coupled with the amplitude \cite{EVG92,TS99}.
Multiplicative noise is the addition of a term that consists of a
noise factor multiplying the amplitude \cite{TNS02}. Since it is
standard to have the driving represented by an appropriate
dimensionless parameter times the amplitude, multiplicative noise
enters the equations as an additive factor to the driving term. A
surprising feature of the theoretical studies is the finding that
the rate of domain growth increases when the noise is sufficiently
small \cite{EVG92,TS99,TNS02}. Since these two very different
noise sources have the same impact,  it is possible that any
generic noise would increase the rate of coarsening.

The details of the experimental apparatus are described in
Ref.~\cite{D00}. The nematic liquid crystal N4 was doped with 0.1
wt\% of tetra n-butylammonium bromide [${\rm(C_4H_9)4N Br}$].
Commercial cells \cite{EHCO} with a quoted thickness of 25~$\mu$m
and 1~cm$^2$ electrodes were used, giving an aspect ratio of 400.
The cells were treated so that the director has a uniform planar
alignment (parallel to the top and bottom of the plates). The
direction of the undistorted director is taken as the x-axis and
the direction perpendicular to the plates is taken as the z-axis.
The y-axis is chosen to form a standard right-handed coordinate
system. The average wavelength of the rolls was $51\ {\rm \mu m}$.
The sample temperature was maintained at $35.0 \pm 0.002\
{\rm^{\circ}C}$.

\begin{figure} [htb]
\includegraphics[width=3.2in]{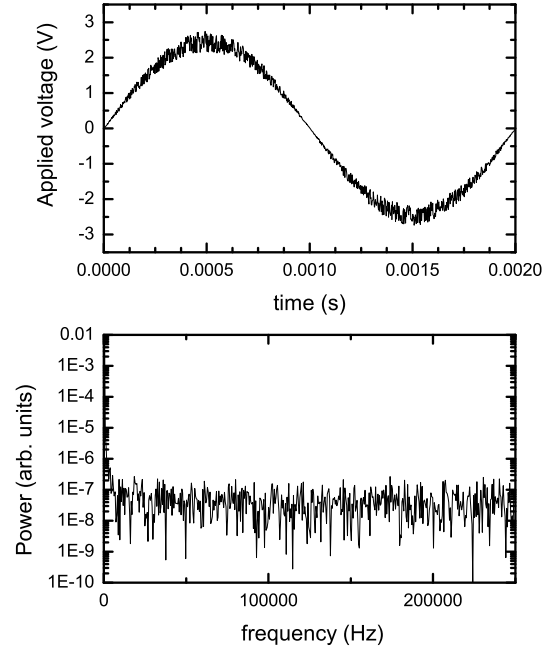}
\caption{(a) A single cycle of a typical waveform that is used to
drive electroconvection. A noise amplitude of 0.6 V is used so
that it is visible to the reader. (b) Power spectrum for the
signal plotted in (a).}
\end{figure}

Typically, an AC voltage of the form $V(t) = V_o\cos(\omega t)$ is
applied perpendicular to the cell, where $V_o$ is the amplitude of
the applied voltage and $\omega/2\pi$ is the driving frequency.
For all of the experiments reported here, $V(t) = [V_o +
\xi(t)]\cos(\omega t)$, where $\xi(t)$ is a random noise term
chosen with a uniform probability from the range $-\xi_m \leq
\xi(t) \leq \xi_m$. Figure~1 shows sample waveforms with their
corresponding power spectra with $\xi_m = 0.6\ {\rm V}$. The
basically flat power spectra indicate the randomness of the noise.
This was achieved by using a built-in pseudo-random number
generator with a seed value that changes often, i.e. the time
function which is the number of seconds elapsed since New Year
1970.  After one cycle of the waveform is randomized, that cycle
is the repeated to create a periodic random waveform.

It should be noted that the noise we add is similar to the
multiplicative noise studied theoretically. However, two facts
need to be considered when comparing our results to theory. First,
we add the noise to the raw voltage and the control parameter is
the square of the voltage. Second, because of experimental
limitations, the noise has a periodic element as only a single
cycle is generated randomly. Therefore, these experiments are
focused on testing the generality of the result that noise
increases the rate of growth of domains, and not on testing a
specific class of noise, such as additive or multiplicative.

\begin{figure} [htb]
\includegraphics[width=3.2in]{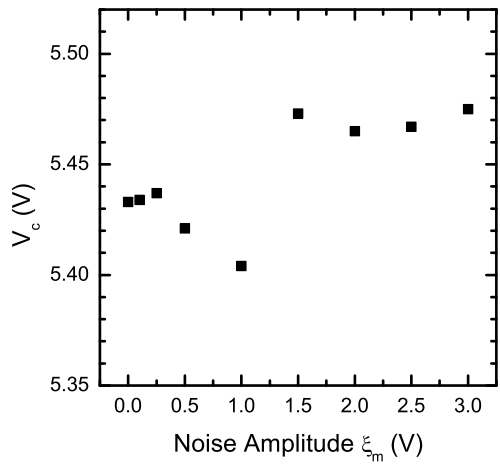}
\caption{Plot of the critical voltage for the onset of
electroconvection as a function of the noise amplitude. Notice,
the transition at a noise amplitude of approximately 1.5 V.}
\end{figure}

The dimensional parameter $\epsilon = (V/V_c)^2 - 1$ characterizes
the depth of the quench. Here $V_c \approx 5.4\ {\rm V}$ is the
onset voltage for electroconvection at the applied frequency ($f =
\omega/2\pi$) of interest. For all of the experiments reported
here, $f = 500\ {\rm Hz}$. The introduction of noise did have a
small effect on the onset voltage, as shown in Fig. 2. In order to
study domain coarsening, voltage quenches were used that started
from a uniform state and suddenly applied sufficient voltage to
reach $\epsilon = 0.2$, where $\epsilon$ is computed using $V_c$
in the absence of noise. This represents one particular type of
quench to compare. The other option would be to have each quench
be to the same value of $\epsilon$ relative to $V_c$ in the
presence of the noise. However, the difference between these two
choices is less than a few percent. At that level, we found the
quench depth did not impact the domain ordering. We also define a
relative noise strength $\eta = \xi_m/V_c$, where $V_c$ is again
taken as the critical voltage in the absence of noise.

After the application of the quench, domains of zig and zag,
separated by grain boundaries and vertical walls of dislocations,
formed in approximately 30 seconds.  As these domains evolved,
images were taken every 30 seconds for 35 minutes. At 35 minutes,
the system essentially is always a single domain within the field
of view. For each noise amplitude, the results of twenty quenches
were averaged. Noise amplitudes ranging from $\eta = 0$ to $\eta =
0.2$ were used.

\begin{figure} [htb]
\includegraphics[width=3.2in]{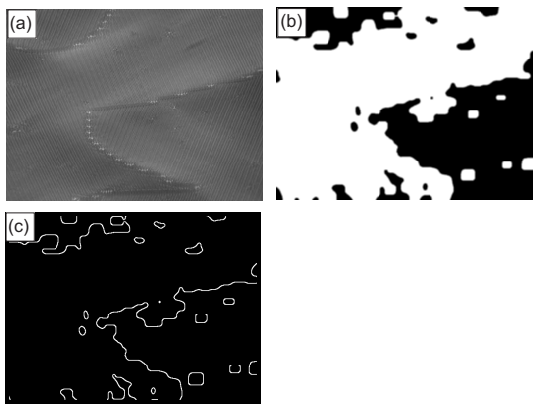}
\caption{Three images illustrating the process by which domains
are defined. The images are all $2.03\ {\rm mm} \times 1.41\ {\rm
mm}$(a) Raw image of electroconvection showing two regions of zag
rolls in between two regions of zig rolls. (b) The processed image
in which zag is white and zig is black. (c) The extracted domain
walls from image (b).}
\end{figure}

\begin{figure} [htb]
\includegraphics[width=3.2in]{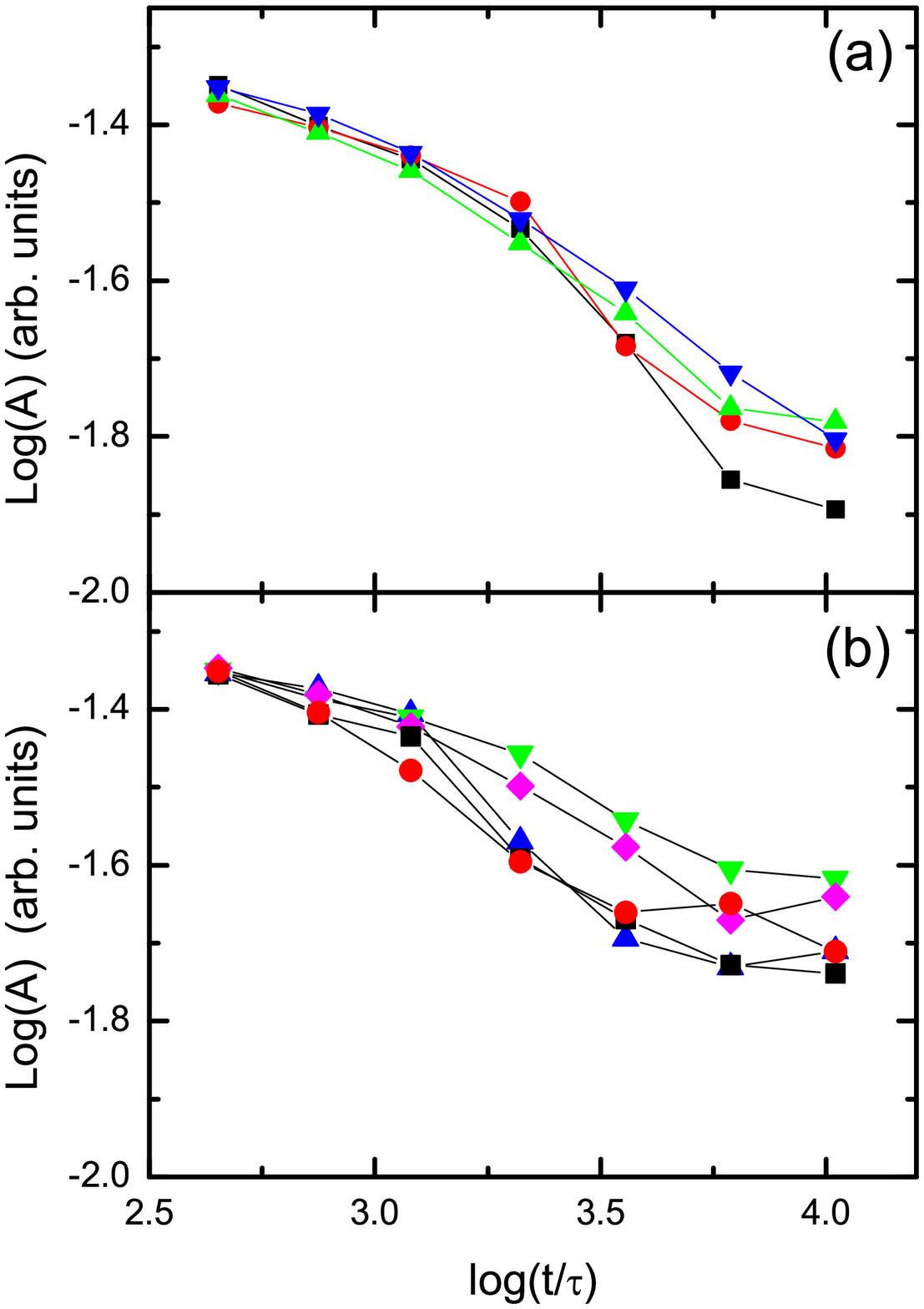}
\caption{(color online)Both plots give the normalized boundary
area as a function of time normalized by the director relaxation
time. (a) is for noise strengths $\eta = 0$ ($\blacksquare$),
0.019 (red $\bullet$), 0.037 (green $\blacktriangle$), 0.056 (blue
$\blacktriangledown$) and (b) is for noise strengths 0.074
($\blacksquare$), 0.093(red $\bullet$), 0.148 (blue
$\blacktriangle$), 0.167 (green $\blacktriangledown$), and 0.185
(magenta $\blacklozenge$). There is a qualitative difference in
the temporal evolution in the two regimes, suggesting a transition
in the temporal evolution as a function of the noise strength.}
\end{figure}

In order to measure the degree of domain coarsening that has taken
place, we look at the length of the boundaries between domains. To
do this, a program was developed using IDL 6.1.  This program
takes advantage of the fact that the product $k_x k_y$ for the
wavevector of the zig pattern is positive while for zag it is
negative. The optical image is translated into a representation of
the sign of $k_x k_y$, with regions of black and white
corresponding to plus and minus, or zig and zag.  The program then
uses thresholding techniques to isolate the boundaries between
domains that are mixtures of black and white pixels.  The number
of boundary pixels represents a way to measure how far the domain
coarsening has progressed. This method is based on the algorithm
described in Ref.~\cite{EMB98}, and the details as applied to our
system are given in Ref.~\cite{PD01}. Figure~3 shows the typical
results of the program for an image that is taken 6 minutes after
the quench.

The main results of the paper are illustrated in Fig.~4. Here the
time evolution of the average domain wall length is plotted for a
number of different noise strengths. Time is measured from
immediately after the quench. Also, time is scaled by the director
relaxation time, $\tau \equiv \gamma_1 d^2/(\pi^2 K_{11}) = 0.2\
{\rm s}$. Here $\gamma_1$ is a rotational viscosity; $K_{11}$ is
the splay elastic constant; and $d$ is the thickness of the cell.
There are two main results.

Figure~4 illustrates that the evolution is slowed by increasing
the noise strength. At a noise strength of 0.07, there is a
qualitative change in the evolution of the system. Below 0.07, the
system appears to be continuing to evolve on the time scale for
which observation was possible. Above 0.07, the system plateaus at
the the later times and the evolution is ``frozen''. Another way
to view this is that the late time measure of domain wall length
is significantly larger for quenches with a noise strength above
0.07, implying that the coarsening dynamics have slowed down. This
is opposite previous theoretical results \cite{EVG92,TS99,TNS02}.
The most likely explanation is the fact that we are using noise
that is not purely multiplicative or additive, but the details of
why this type of noise slows the dynamics requires further study.

Another interesting result of applying this type of noise to the
system is the global impact on the pattern. Under sufficiently
strong amplitude, one would expect the noise to cause the
spontaneous generation of dislocation pairs, resulting in some
level of chaotic dynamics. We observed no evidence for such a
transition, even at noise strengths as high as 0.9. This behavior
has interesting implications for the nature of coupling between
the applied noise and the pattern dynamics. The noise clearly
impacted the dynamics of the system by significantly slowing
domain growth at a critical value. However, the lack of any
generation of defects is probably connected to the fact that the
noise is applied directly to the voltage and that the driving
force is given by the root mean square value of the voltage. The
impact of this should be explored further with simulations and by
experimental studies of different noise sources. It also presents
an interesting pattern situation in which the dynamics of a system
can be controlled to some degree by noise without having a
substantial impact on the qualitative aspects of the pattern.

\begin{acknowledgments}

This work was supported by Department of Energy grant
DE-FG02-03ED46071.

\end{acknowledgments}


\end{document}